\documentclass[11pt]{article}
\usepackage{moriond,epsfig,amsmath}

\def\Journal#1#2#3#4{{#1} {\bf #2}, #3 (#4)}

\def\NPB{{\em Nucl. Phys.} B}
\def\PLB{{\em Phys. Lett.}  B}
\def\PRL{\em Phys. Rev. Lett.}
\def\PRD{{\em Phys. Rev.} D}

\def\epp{\epsilon^{\prime}}
\def\ree{\ensuremath{Re(\epp/\epsilon)}}

\def\ra{\ensuremath{\rightarrow}}

\def\kl{\ensuremath{K_L}}
\def\ks{\ensuremath{K^{\ast}}}
\def\gs{\ensuremath{\gamma^{\ast}}}
\def\ee{\ensuremath{e^+e^-}}
\def\eeg{\ensuremath{\ee\gamma}}
\def\mm{\ensuremath{\mu^+\mu^-}}
\def\pp{\ensuremath{\pi^+\pi^-}}
\def\ppg{\ensuremath{\pp\gamma}}
\def\pz{\ensuremath{\pi^0}}
\def\pzd{\ensuremath{\pi^0_D}}
\def\pzpz{\ensuremath{\pz\pz}}
\def\pzpzg{\ensuremath{\pzpz\gamma}}

\def\ktpd{\ensuremath{\kl\ra\pz\pz\pzd}}
\def\kpp{\ensuremath{\kl\ra\pp}}

\def\kpmzd{\ensuremath{\kl\ra\pp\pzd}}
\def\kppg{\ensuremath{\kl\ra\ppg}}
\def\kppee{\ensuremath{\kl\ra\pp\ee}}
\def\kpzpzg{\ensuremath{\kl\ra\pzpzg}}
\def\kpee{\ensuremath{\kl\ra\pz\ee}}
\def\klt{\ensuremath{\kl\ra\pi^{\pm}l^{\mp}\nu}}

\def\ketg{\ensuremath{\kl\ra\pi^{\pm}e^{\mp}\nu\gamma}}
\def\ketee{\ensuremath{\kl\ra\pi^{\pm}e^{\mp}\nu\ee}}
\def\pee{\ensuremath{\pz\ra\ee}}
\def\peeg{\ensuremath{\pz\ra\ee\gamma}}
\def\pdeeg{\ensuremath{\pzd\ra\ee\gamma}}
\def\pgg{\ensuremath{\pz\ra\gamma\gamma}}
\def\keeg{\ensuremath{\kl\ra\eeg}}
\def\kmm{\ensuremath{\kl\ra\mm}}
\def\kgsg{\ensuremath{\kl\gs\gamma}}
\def\kgsgs{\ensuremath{\kl\gs\gs}}
\def\order{\ensuremath{\cal{O}}}
\def\adip{\ensuremath{\alpha_{DIP}}}
\def\aks{\ensuremath{\alpha_{\ks}}}

\begin{document}
\vspace*{4cm}
\title{RECENT RESULTS FROM THE KTEV EXPERIMENT}

\author{ M. J. WILKING }

\address{University of Colorado.  Department of Physics.\\
390 UCB, Boulder, CO  80302, USA}

\maketitle\abstracts{ We present recent preliminary results from five
  decay channels.  From the \kppg\ channel, we extract form factors
  for the CP violating M1 direct photon emission amplitude and the
  fraction of the total decay amplitude that is due to direct
  emission.  We have placed an upper limit on the \kpzpzg\ branching
  ratio, and preliminary measurements of the \ketee\ and \pee\
  branching ratios are presented.  Finally, we report measurements of
  both the branching ratio and the form factor parameters for the
  decay \keeg.
}

\section{The KTeV Detector}

The KTeV detector was used by two Fermilab experiments.  The first,
E799-II, was designed to measure rare kaon decays such as \kpee\ and
\kppee.  The second, E832, was designed to measure the CP violating
parameter \ree.

The two neutral kaon beams used by KTeV passed through a 60m vacuum
decay volume.  Downstream of the decay volume was a charged
spectrometer, which consisted of four drift chambers and a dipole
magnet.  The spectrometer determined charged particle momenta with a
resolution of $\sigma(P)/P = 0.38\%\oplus0.16\%*P$(GeV/c).  Beyond the
spectrometer was a transition radiation detector (TRD), which was used
to separate electrons and pions.  Immediately following the TRD was a
3100 crystal CsI calorimeter.  The energy resolution achieved on
clusters in the calorimeter was $\sigma(E)/E = 0.45\%\oplus
2\%/\sqrt{E(\textrm{GeV})}$.  Behind the calorimeter were alternating
layers of steel and scintillator used to reject backgrounds with muons
in the final state.

\section{The Decay \kppg}
\subsection{Motivation}\label{subsec:kppgmot}

The decay \kppg\ can proceed through two channels.  The inner
Bremsstrahlung (IB) channel is identical to the CP violating \kpp\
decay, except that an additional photon is internally radiated by one
of the charged pions.  In the direct emission (DE) channel, all three
decay particles emerge directly from the \kl\ decay vertex.  The DE
process usually occurs via a CP conserving M1 term.  However, there
can also be a small contribution from a CP violating E1 term.  The E1
DE amplitude can contribute via interference with the IB amplitude as
well.

The expressions for each of these decay amplitudes are given in Eqs.
\ref{eq:ib}-\ref{eq:ede}.  Using these expressions, the values of the
parameters $|\tilde{g}_{M1}|$, $a_1/a_2$, and $|g_{E1}|$ can be
extracted from the measured amplitudes.
\begin{eqnarray}
E_{IB}(\kl) & = & \left(2\frac{M_K}{E_{\gamma}}\right)^2
     \frac{|\eta_{+-}|e^{i\Phi_{+-}}e^{i\delta_0}}
     {1-\left(1-\frac{4m_{\pi}^2}{M_{\pi\pi}^2}\right)
     \cos^2(\theta)} \label{eq:ib} \\
M_{DE}(\kl) & = & i\tilde{g}_{M1}
     \left(1+\frac{a_1/a_2}{(M_{\rho}^2-M_K^2)+2M_KE_{\gamma}}\right)
     e^{i\delta_1} \label{eq:mde} \\
E_{DE}(\kl) & = & |g_{E1}|e^{i\delta_1} \label{eq:ede}
\end{eqnarray}
In the above expressions, $M_K$, $M_{\rho}$, and $M_{\pp}$
are the kaon mass, rho mass, and the mass of the \pp\ system,
respectively.  $E_{\gamma}$ is the photon energy in the kaon rest
frame and $\theta$ is the angle between the $\gamma$ and $\pi^+$ in
the \pp rest frame.  $\delta_0(s=M_K^2)$ and
$\delta_1(s=M_{\pi\pi}^2)$ are the isospin 0 and 1 strong phase
shifts, and $|\eta_{+-}|e^{i\Phi_{+-}}$ is the \kpp\ amplitude.

\subsection{\kppg\ Analysis}

\begin{figure}[h]
  \begin{minipage}[h]{.45\textwidth}
    \begin{center}
    \epsfig{figure=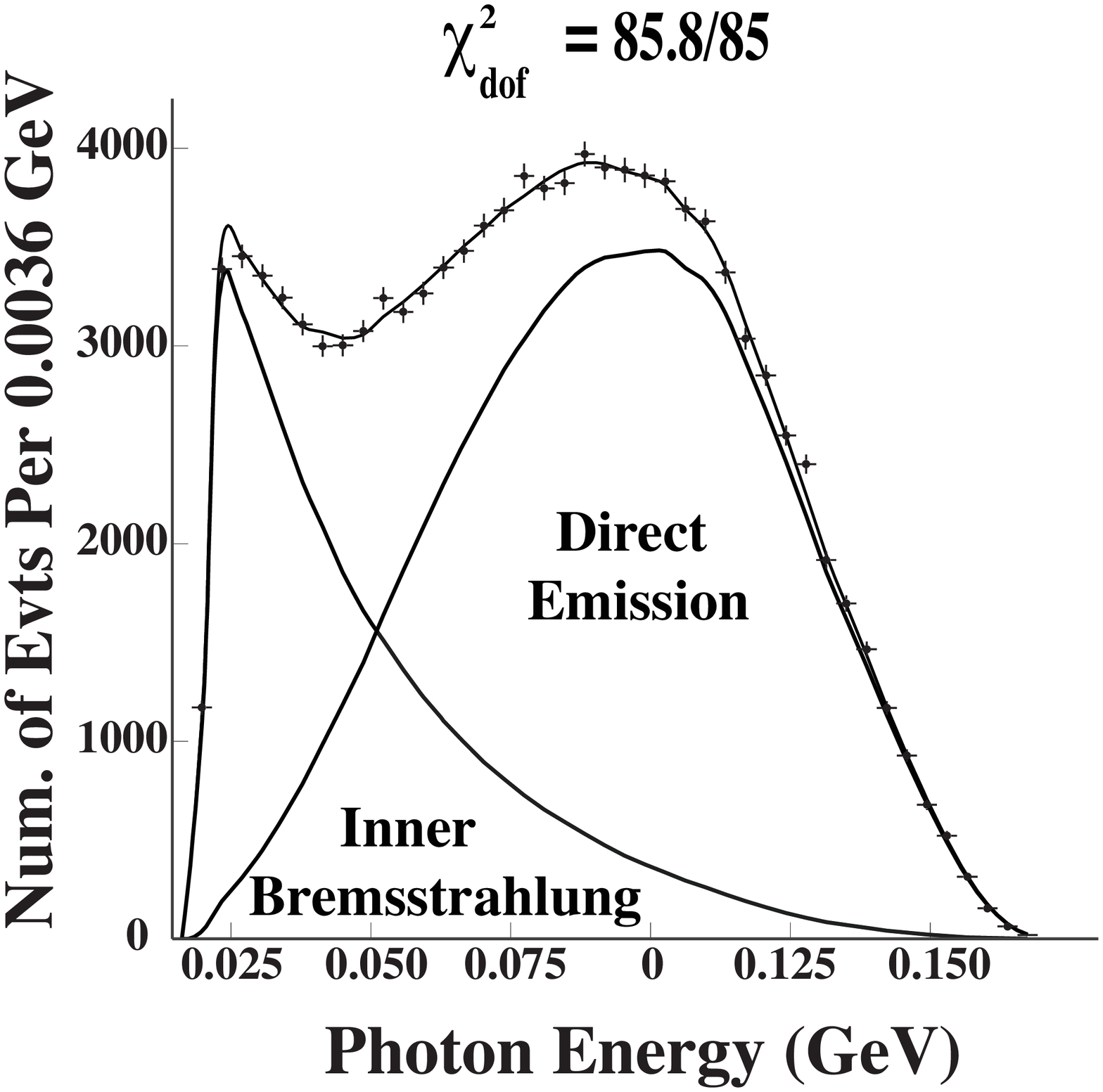,height=5cm}
    \caption{The Inner Bremsstrahlung and Direct Emission components of
             the $E_{\gamma}$ distribution are shown along with the data
             (points) and the total Monte Carlo (curve).
    \label{fig:egam}}
    \end{center}
  \end{minipage}
  \hfill
  \begin{minipage}[h]{.45\textwidth}
    \begin{center}
    \epsfig{figure=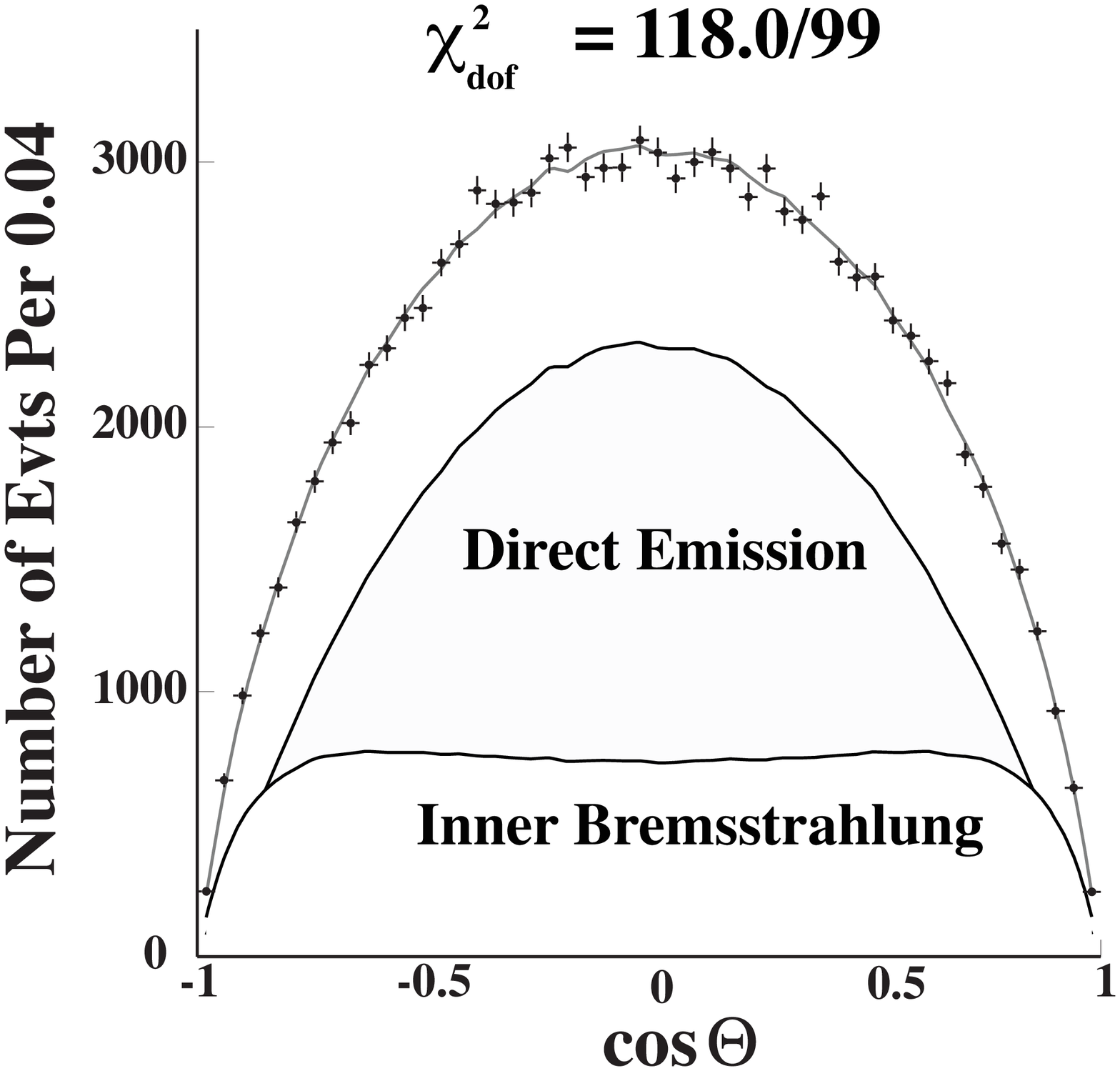,height=5cm}
    \caption{The Inner Bremsstrahlung and Direct Emission components of
             the $\cos(\theta)$ distribution are shown along with the data
             (points) and the total Monte Carlo (curve).
    \label{fig:cost}}
    \end{center}
  \end{minipage}
\end{figure}

The IB and M1 DE decay amplitudes can be separated in the $E_{\gamma}$
and $\cos(\theta)$ variables described in Sec.
\ref{subsec:kppgmot}.  The Monte Carlo prediction for both of these
components is compared to data as shown in Figs. \ref{fig:egam} and
\ref{fig:cost}.  A two-dimensional likelihood fit is performed in
$E_{\gamma}$ and $\cos(\theta)$ to extract the values of
$|\tilde{g}_{M1}|$ and $a_1/a_2$.  Despite isolating $111.4\times10^3$
signal events over a $0.4\%$ background, the E1 DE amplitude is still
too small to be measured.

The results for \kppg\ have been submitted to {\PRL}\cite{ppgpre}.
The fit results for the M1 direct emission parameters are
\begin{eqnarray}
|\tilde{g}_{M1}| & = & 1.198 \pm 0.035(stat) \pm 0.086(syst)
     \label{eq:gm1fit} \\
a_1/a_2 & = & -0.738 \pm 0.007(stat) \pm 0.018(syst)
     \textrm{GeV}^2/\textrm{c}^2
     \label{eq:a1a2fit}
\end{eqnarray}
These measurements are the most precise to date.  We also measure the
direct emission fraction and an upper limit on the CP violating E1
direct emission parameter $g_{E1}$:
\begin{eqnarray}
DE / (DE+IB) & = & 0.689 \pm 0.021~~~(E_{\gamma}>20 MeV) \\
g_{E1} & < & 0.21~~~\textrm{(90\% confidence)}
\label{eq:ge1fit}
\end{eqnarray}

\section{The Decay \kpzpzg}
\subsection{Motivation}

The decay \kpzpzg\ can only proceed via direct emission from the
\kl\pzpz\ vertex.  Since the two pions in the decay are identical, the
lowest contributing multipole moment is L=2.  The E2 direct emission
term is CP conserving, while the M2 direct emission term is CP
violating.

The decay amplitude for this mode vanishes to \order$(p^4)$ in chiral
perturbation theory (ChPT), and therefore provides a direct probe of
\order$(p^6)$ ChPT.  The predicted branching ratio from ChPT is
$(7\times10^{-11})$\cite{ecker}.  However, extrapolations from the
\kppg\ branching ratio can give predictions as high as
$(1\times10^{-8})$\cite{heiliger}.  The current upper limit from the NA31
experiment is $(5.6\times10^{-6})$\cite{barr}.

\subsection{\kpzpzg\ Analysis}

The search for \kpzpzg\ is performed using a blind analysis.  We
isolate $\kl\pz\pzd\gamma$ $(\pgg,\pdeeg)$ to utilize our sensitive
charged particle trigger.  There is a large background due to \ktpd\
events with a missing photon, however most of these events can be
eliminated by requiring that the \ee\ mass lie within 3 MeV/c$^2$ of
the kaon mass and that the square of the momentum transverse to the
kaon direction be less than 0.00015 GeV$^2$/c$^2$.  The analysis cuts
retain 80\% of the signal Monte Carlo.

The preliminary result presented here is based on 40\% of the total
data set.  The Monte Carlo gives a prediction of $1.66\pm0.59$
background events in the signal region.  No events were found in the
signal region.  We set an upper limit on the branching ratio value of
\begin{equation}
BR(\kpzpzg) < 2.52\times10^{-7}~~~\textrm{(90\% confidence)}
\end{equation}
This measurement represents a factor of 22 improvement on
the NA31 result.

\section{The Decay \ketee}
\subsection{Motivation}

The CKM matrix element $V_{us}$ can be extracted from \klt\ decays via
\begin{equation}
\Gamma_{\klt} = \frac{G_F^2M_K^5}{192\pi^3}S_{EW}
     \left(1+\delta_K^l\right)C^2|V_{us}|^2f_+^2(0)I_K^l
\label{eq:gklt}
\end{equation}
where $l$ refers to either $e$ or $\mu$, $G_F$ is the Fermi
constant, $M_K$ is the kaon mass, $S_{EW}$ is the short-distance
radiative correction, $\delta_K^l$ is the long-distance radiative
correction, $f_+(0)$ is the form factor at zero momentum transfer to
the $l\nu$ system, and $I_K^l$ is the phase-space integral, which
depends on the measured semileptonic form factors.  The value of $C^2$
is 1 for neutral kaon decays and 1/2 for charged kaon decays.

The uncertainty in the $K\pi W$ vertex parameters is currently larger
than the experimental uncertainty in $\Gamma_{\klt}$.  The direct
emission \ketg\ decay is sensitive to the structure of the $K\pi W$
vertex, but unfortunately this decay is dominated by inner
Bremsstrahlung.  If the radiated photon in \ketg\ is virtual, there is
an enhancement in the DE/IB ratio, therefore the decay \ketee\ is a
more sensitive probe of the $K\pi W$ vertex.

\subsection{\ketee\ Analysis}

The main background to \ketee\ decays comes from \ktpd\
($\pzd\ra\ee\gamma$) where the photon is missing and one of the pions
is mis-reconstructed as an electron.  This background can be greatly
reduced by cutting on the square of the longitudinal momentum of the
missing \pz\ in the frame where the kaon momentum is perpendicular to
the \pp\ momentum.  If this quantity is negative, it indicates that
there is not enough phase space to form a \pz\ if the missing energy
is assumed to be a photon.

We have isolated 19466 signal events with a background of 4.95\%.  To
measure the absolute branching ratio, the \ketee\ rate is normalized
to the \kpmzd\ (\pdeeg) rate.  We find 300526 normalization mode
events with a background of 2.23\%.  In addition, we define \ketee\
events to have an \ee\ mass greater that 5 MeV/c$^2$.  The preliminary
branching ratio measurement is
\begin{equation}
\begin{split}
BR(\ketee, M_{\ee}>5\textrm{MeV/c}^2)
     = (1.606 & \pm 0.012(stat)~^{+0.026}_{-0.016}(syst) \\
     & \pm 0.045(ext.syst))\times 10^{-5}
\label{eq:brketee}
\end{split}
\end{equation}
The additional $(ext.syst)$ term is due to the experimental
uncertainty on the branching ratio of \kpmzd.

\section{The Decay \pee}
\subsection{Motivation}

The decay \pee\ proceeds through a loop process at lowest order.  This
mode provides a sensitive probe of theoretical models due to the
precision with which it can be measured and the seemingly simple
nature of the process.  The contribution from on-shell photons sets a
lower unitary bound on the rate:
\begin{equation}
\Gamma(\pee)/\Gamma(\pgg) \ge 4.75\times 10^{-8}
\end{equation}
Any additional contribution from off-shell photons will enhance the
rate.  Vector meson dominance (VMD) models predict a branching ratio
of $6 - 6.4\times 10^{-8}$\cite{bergstrom}\cite{ametller}, whereas
predictions from chiral perturbation theory have a range of
$6.2-8.3\times 10^{-8}$\cite{knecht}\cite{gomez}.

\subsection{The \pee\ Measurement}

\begin{figure}[h]
 \begin{center}
 \epsfig{figure=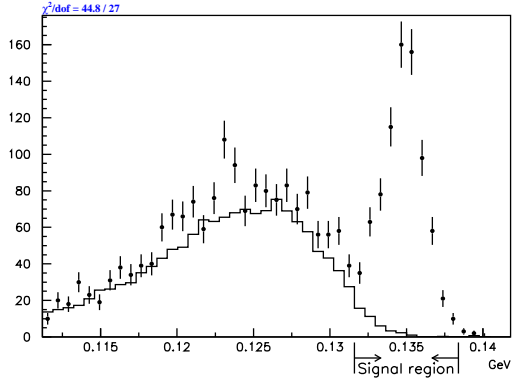,height=5cm}
 \caption{The \ee\ mass spectrum for \pee\ events is shown.  The points with error bars are the data and the histogram is the background Monte Carlo only.
 \label{fig:peemass}}
 \end{center}
 \end{figure}

The main background to \pee\ is the decay \peeg\ where the photon is
either lost or very soft.  Fortunately, the \peeg\ \ee\ mass spectrum
falls off very quickly as the \ee\ mass approaches the \pz\ mass,
whereas the \pee\ spectrum peaks sharply at high \ee\ mass.
Figure
\ref{fig:peemass} shows the \ee\ mass spectrum in data compared to the
Monte Carlo background estimate.  A clear \pee\ peak can be seen at
the pion mass.
The main systematic uncertainty comes from the
mismatch between the background level predicted by the Monte Carlo and
the amount of background observed in the data.

There are 714 events in the signal region with a predicted background
of $39.9\pm 12.3$ events.  The \pee\ rate is normalized to the \peeg\
rate.  The preliminary result for this ratio is
\begin{equation}
\frac{BR(\pee,x>0.95)}{BR(\peeg,x>0.232)}
     = (1.721 \pm 0.068(stat) \pm 0.036(syst)) \times 10^{-4}
\end{equation}
where $x$ is the squared ratio of the \ee\ mass to
the \pz\ mass.  Using the \peeg\ branching ratio and the full \ee\
spectrum, we find
\begin{equation}
BR(\pee,x>0.95) = (6.56 \pm 0.26(stat) \pm 0.23(syst)) \times 10^{-8}
\end{equation}
The systematic error now includes the 2.7\% uncertainty in
the \peeg\ branching ratio and the 0.5\% uncertainty in the \pz\ slope
parameter.

\section{The Decay \keeg}
\subsection{Motivation}

The decay \keeg\ is interesting due to the implications it has on the
interpretation of the \kmm\ rate.  The decay \kmm\ contains
short-distance contributions from which one can extract $|V_{td}|$.
However, the \kmm\ rate is dominated by long distance contributions
containing a \kgsgs\ vertex, which must first be subtracted.  The
\kgsgs\ vertex can be probed by various double and single Dalitz
decays such as \keeg.

\subsection{\kgsg\ Form Factor}

Two form factor models were considered in this analysis. The
D'Ambrosio, Isidori, and Portoles (DIP) model is a general two term
model consistent with \order$(p^6)$ chiral perturbation theory.  The
\keeg\ process is sensitive to the parameter \adip.  Bergstrom, Masso,
and Singer (BMS) have proposed a vector meson dominance model that
depends on the parameter \aks\ as shown:
\begin{equation}
f_{BMS}(x)=\frac{1}{1-x\frac{M_K^2}{M_{\rho}^2}}
     +\frac{C~\aks}{1-x\frac{M_K^2}{M_{\ks}^2}}
     \left(\frac{4}{3}-\frac{1}{1-x\frac{M_K^2}{M_{\rho}^2}}
     -\frac{1}{9}\frac{1}{1-x\frac{M_K^2}{M_{\omega}^2}}
     -\frac{2}{9}\frac{1}{1-x\frac{M_K^2}{M_{\phi}^2}}\right)
\label{eq:bms}
\end{equation}
In this context, the variable $x$ is the squared ratio of the \ee\
mass to the kaon mass.  The $M_i$ variables correspond to the mass of
the indicated meson, and the constant $C$ is given by the following
expression.
\begin{equation}
C=(8\pi\alpha_{EM})^{1/2}G_{NL}f_{\ks K\gamma}m_{\rho}^2/(f_{\ks}f_{\rho}^2A_{\gamma\gamma})
\label{eq:cexp}
\end{equation}
The value of $C$ used in past measurements has not been
consistent despite the fact that measurements of \aks\ are often
compared directly.  In some cases, $C$ is recalculated using the most
recent values for the physical inputs to Eq. \ref{eq:cexp}, and in
others, the value of $C$ from a previous publication is used.  To
avoid the issue of a time varying value for $C$, we fit for the
combined quantity $C\aks$.

\begin{figure}[h]
\begin{center}
\epsfig{figure=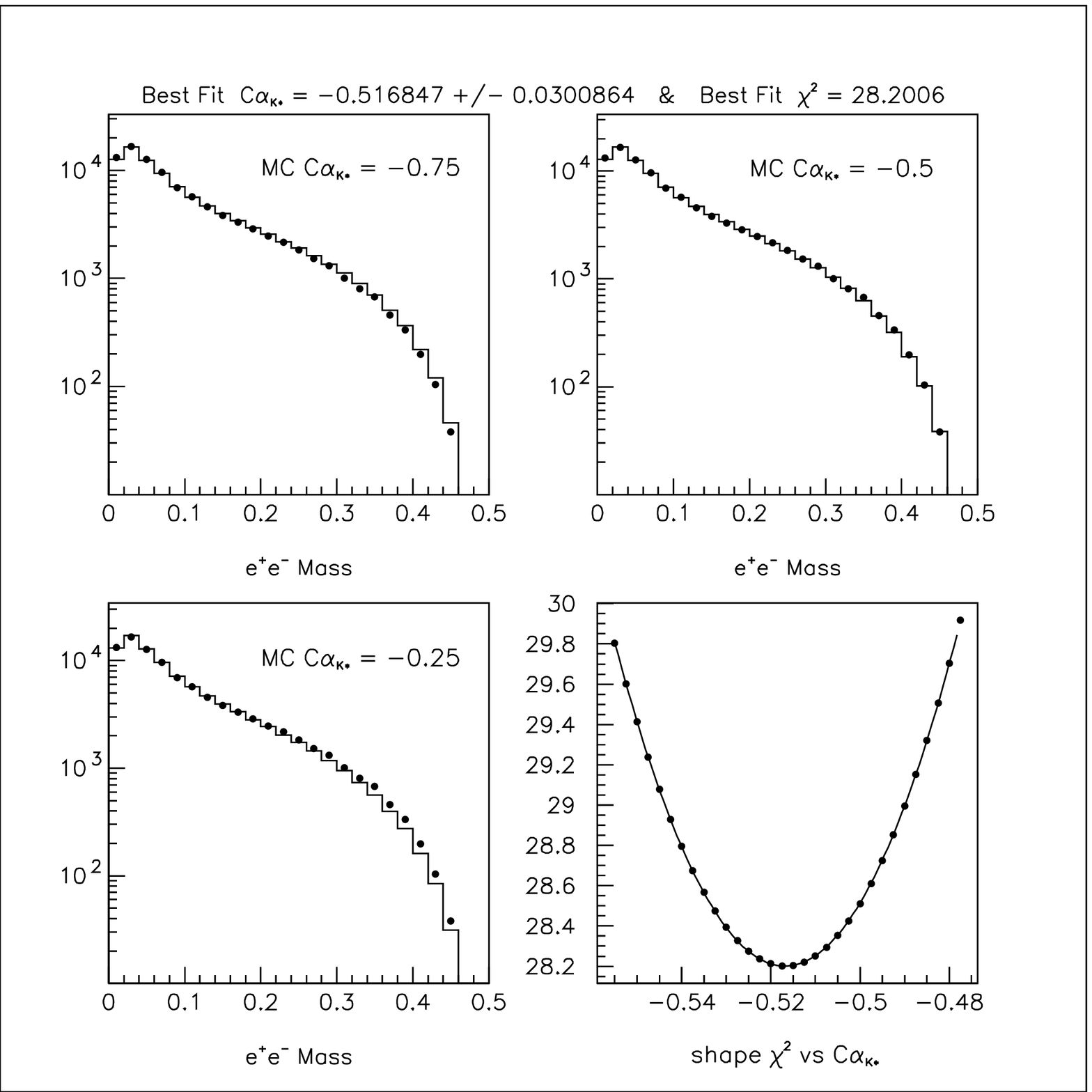,height=7cm,clip=true}
\caption{The \keeg\ data (dots) is compared to three sets of Monte Carlo, each with a different value of $C\aks$.  The lower right shows the fit to the shape-$\chi^2$ versus $C\aks$ distribution.
\label{fig:keegff}}
\end{center}
\end{figure}

The shape of the \ee\ mass distribution is very sensitive to the form
factor.  To extract a measurement of the form factor parameter, a
bin-by-bin shape-$\chi^2$ fit is performed between the data and sets
of Monte Carlo with differing values of $\alpha$.  Figure
\ref{fig:keegff} shows three comparisons of data to Monte Carlo and
the fit to the shape-$\chi^2$ versus $C\aks$ distribution.

\subsection{Preliminary Results}

The measured branching ratio for \keeg\ is
\begin{equation}
BR(\keeg) = (9.25\pm0.03(stat)\pm0.07(syst)\pm0.26(ext.syst))\times 10^{-6}
\end{equation}
The $(ext.syst)$ term is due to the uncertainty in the branching ratio
of the normalization mode, \ktpd.  The form factor fits yield the
following values:
\begin{eqnarray}
C\aks & = & -0.517 \pm 0.030(fit) \pm 0.022(syst) \\
\adip & = & -1.729 \pm 0.043(fit) \pm 0.028(syst)
\end{eqnarray}

\end{document}